\documentclass[11pt]{article}
\usepackage{amsfonts}
\usepackage{amsmath,amssymb}

\newcommand{\pa}{\partial}
\newcommand{\n}{\nonumber\\}
\newcommand{\om}{\Omega}

\newcommand{\bea}{\begin{array}}
\newcommand{\ear}{\end{array}}
\newcommand{\noi}{\noindent}
\newcommand{\me}{\frac{1}{2}}
\newcommand{\uu}{{\bf u}}
\newcommand{\RR}{\mathbb{R}}
\newcommand{\ap}{\alpha}

\newcommand{\ot}{\otimes}

\newcommand{\bege}{\begin{equation}}
\newcommand{\enge}{\end{equation}}

\newcommand{\ri}{\rightarrow}
\newcommand{\si}{\sigma}

\newcommand{\beq}{\begin{eqnarray}}
\newcommand{\eeq}{\end{eqnarray}}

\newcommand{\EE}{\mathbb{E}}

\newcommand{\vbn}{\blacktriangleleft}
\newcommand{\vvn}{\blacktriangleright}

\newcommand{\mr}{\mathring}

\usepackage{indentfirst}

\begin{document}
\title{AdS geometry, projective embedded coordinates and associated isometry groups}
\author{{\bf Rold\~ao da Rocha}\thanks{DRCC - Instituto de F\'{\i}sica Gleb Wataghin, 
 Unicamp, CP6165,  Campinas, SP, Brazil. 
13083-970.
E-mail: {\tt roldao@ifi.unicamp.br.} Supported by CAPES.}\and
{\bf E. Capelas de Oliveira }\thanks{Departamento de Matemática Aplicada, IMECC, Unicamp, CP 6065, 13083-859, Campinas, SP, Brazil. E-mail: 
{\tt capelas@ime.unicamp.br}}}
\date{}\maketitle \bigskip
 \abstract
${}^{}$
\begin{center}
\begin{minipage}{13,5cm}$ \;\;\;\;\;$ \scriptsize{This work is intended to investigate the geometry of anti-de Sitter spacetime (AdS), 
from the point of view of the Laplacian Comparison Theorem (LCT), and to give another description of
the hyperbolical embedding standard formalism of the de Sitter and anti-de Sitter spacetimes in a pseudo-Euclidean 
spacetime.  After Witten proved that general relativity is a renormalizable quantum system in (1+2) dimensions,
it is possible to point out few interesting motivations to investigate AdS spacetime. A lot of attempts were made to generalize
the gauge theory of gravity in (1+2) dimensions to higher ones. The first one was to enlarge the Poincar\'e group of symmetries,
 supposing 
an AdS group symmetry, which contains the Poincar\'e group. Also, the AdS/CFT correspondence
 asserts that a  maximal supersymmetric Yang-Mills theory in 4-dimensional Minkowski spacetime is equivalent to a type IIB closed 
superstring theory. The 10-dimensional arena for the type IIB superstring theory is described by the product manifold 
$S^5\times$ AdS, 
an impressive consequence that motivates the investigations about 
the AdS spacetime in this paper, together with the de Sitter spacetime. 
Classical results in this mathematical formulation are reviewed in a 
more general setting together with the isometry group associated to the de Sitter spacetime. 
It is known that, out of the Friedmann models that describe our universe, the Minkowski, de Sitter and anti-de Sitter
spacetimes are the unique maximally isotropic ones, so they admit a
maximal number of conservation laws and also a maximal number of Killing
vectors. In this paper it is shown how to reproduce some geometrical properties of AdS, from 
the LCT in AdS, choosing suitable
functions that satisfy basic properties of Riemannian geometry.
  We also introduce and discuss the well-known embedding of a 4-sphere
and a 4-hyperboloid in a 5-dimensional pseudo-Euclidean spacetime, reviewing the
usual formalism of spherical embedding and the way how it can retrieve the Robertson-Walker metric.
 With the choice of the de Sitter metric static frame, we write the so-called reduced  model in suitable 
coordinates. We assume the existence of projective coordinates, since de Sitter spacetime is orientable.  From these coordinates,
 obtained when stereographic projection of the de Sitter 
4-hemisphere is done, we consider the Beltrami geodesic representation, which gives a more general formulation of the seminal full model
 described by  Schr\"odinger, concerning
 the geometry and the topology of de Sitter spacetime. Our formalism retrieves the classical one  if we consider the 
metric terms over the de Sitter splitting on Minkowski spacetime. From the covariant derivatives we find the acceleration of
 moving particles, Killing vectors and the isometry group generators associated to de the Sitter spacetime.}
\end{minipage}\end{center}
\medbreak
\medbreak \noindent
\medbreak \noindent
Pacs numbers:  02.20.-a, 02.40.-k, 04.50.+h 
\section*{Introduction}
 Today it is of great interest to investigate  
isometry groups of a given universe model, and it becomes natural to ask whether some models
admit the energy conservation law. We restrict our attention to de Sitter (dS) and anti-de Sitter (AdS) spacetimes, which are 
respectively solutions 
of  Einstein equations with cosmological constant $\Lambda = \pm 3/R^2$ ($R>0$), and curvature given by the components 
$R_{\mu\nu} = \Lambda g_{\mu\nu}$ of the Ricci tensor. These manifolds have shown themselves suitable as geometric arenas to
 investigate conformal field 
theories \cite{c1} and superstrings \cite{c2}. Since de Sitter group is the maximal inner group contained in the conformal group, many physical theories
are formulated in dS and AdS scenarios. 
The maximally compact subgroup of SO(3,2), the symmetry group of the dS$_{3,2}$ spacetime, is SO(3)$\times$SO(2), which is two-fold 
covered by SU(2)$\times$U(1). It can therefore be 
used as an alternative formalism to describe the Glashow-Weinberg-Salam model of electroweak interactions \cite{c3}, since
the gauge group  SU(2)$\times$U(1) is related to the isospin and weak hypercharge of elementary particles. 
The group SO(3,2) is also a dynamical group associated to the \emph{Zitterbewegung} \cite{1,c4}. 
 dS and AdS spacetimes
 also allow exact solutions of the field equations and the symmetry group SO(4,1) is used to classify physical states \cite{2,3}.

dS spacetime is geometrically described as a 4-hyperboloid with topology $S^3 \times \RR$. If a Wick rotation
of the time coordinate is performed, dS can be seen as 
a 4-sphere, and then it is  possible, using the methods of projective geometry \cite{4}, to 
describe dS spacetime
in terms of Minkowski orthogonal coordinates.
 From now on dS$_{4,1}$ denotes the de Sitter spacetime embedded in a pseudo-Euclidean spacetime endowed
 with a metric of signature (4,1). 

This article is organized as follows: in Sec. 1   
the metric in  AdS is obtained, describing the geometry of a simply connected hyperbolic spacetime of constant sectional curvature $k$.
The Laplacian Comparison Theorem is also proved, from the viewpoint of AdS geometry, which is briefly reviewed. 
In Sec. 2  we present and discuss the main differences between 
the spherical and hyperbolical embedding in a pseudo-Euclidean spacetime, retrieving the Robertson-Walker 
metric by the introduction of an appropriate expansion parameter. Some 
of the features of the Schr\"odinger model \cite{6}
are reviewed, together with a brief exposition of the de Sitter metric,
 using the static frame.  In Sec. 3, after  the metric of dS$_{4,1}$ is constructed, 
the covariant derivative and subsequently the acceleration of a moving particle are explicitly given in terms 
of the Christoffel connection symbols. The Killing vectors are obtained from the Killing equations, and also the generators of
the isometry group of dS$_{4,1}$. In Sec. 4, 
after performing a Wick rotation in time coordinate, dS$_{4,1}$ spacetime is described 
as a 4-sphere and the Beltrami representation is obtained. It gives an useful way to describe embedded 
projective coordinates as the usual orthogonal coordinates on Minkowski spacetime. The de Sitter metric, which is retrieved if we consider
 the linear metric tensor components on Minkowski spacetime, is generalized. 

\section{Revisiting the AdS geometry and the Laplacian Comparison Theorem} 
In this section we revisit the
approach given in \cite{esc} in the case  when the manifold $M$ is given by the hyperbolic AdS spacetime.
From now on we denote $\pa_\ap = \pa/\pa_\ap$ and we call by {\it metric} a non-degenerate symmetric bilinear form.

Consider a $C^2$ function $\psi: M \ri \RR$ and  vector fields $X,Y\in \Omega(M)$. The Hessian of $\psi$ is defined as
\bege
H\psi(X,Y) = D_X d\psi(Y) = XY(\psi) - D_X Y(\psi).
\enge\noi 
The metric in $\RR^n$ is given by 
\bege\label{mep}
g = dr\ot dr + \psi^2(r) d\Omega\ot d\Omega,
\enge\noi where $d\Omega\ot d\Omega: T_x(S^{n-1})\times  T_x(S^{n-1}) \rightarrow \RR $ is the metric on 
the $(n-1)$-sphere $S^{n-1}$, for $x = r\Omega$, $r>0$, $\Omega\in S^{n-1}$. The tangent space at $x \in S^{n-1}$ is denoted by $T_x(S^{n-1})$.
From the definition of the Hessian, and using the fact that the curves $r\Omega$ for $\Omega$ fixed are geodesics, and $\pa_r$ is the velocity vector, we have
\bege
Hr \left({\pa_r}, {\pa_r}\right) = {\pa_r}{\pa_r} r - D_{{\pa_r}}{\pa_r} r = 0.
\enge\noi Besides, if $X\cdot {\pa_r} = 0$, the expression
\bege
Hr\left({\pa_r}, X\right) = Hr\left(X,{\pa_r}\right) = X\left({\pa_r}r\right) - \left(D_X{\pa_r}\right)r
\enge\noi follows. But since $(D_X{\pa_r})r = (D_X{\pa_r})\cdot {\pa_r}$, if we suppose that the radial geodesics are parametrized
by the arc length, i.e.,
\bege\label{d1}
{\pa_r}\cdot {\pa_r} = 1,\enge we can covariantly differentiate eq.(\ref{d1}) with respect to $X$, obtaining
$(D_X{\pa_r})\cdot {\pa_r} = 0$. Therefore,
\bege\label{part}
Hr\left({\pa_r}, X\right) = 0.
\enge\noi 
Let $a\in\RR_+^*$, and suppose that the vector fields $X$ and $Y$ are tangent to the level hypersurface $r=a$. Then
\bege
Hr(X,Y) = -(D_XY)(r) = -(D_XY)\cdot \pa_r = Y\cdot (D_X\pa_r),
\enge\noi since $Y\cdot\pa_r = 0$. Now, let $\{e_i\}_{i=1}^{n-1}$ be
 a set of orthonormal (coordinate) frame on $S^{n-1}$, and
without loss of generality, let $e_n:= \pa_r$ be 
the normal vector field to the hypersurface $r=c$. On one hand we have
\beq
e_i\cdot (D_{e_j}\pa_r) = e_i\cdot (D_{e_j}e_n) = \Gamma_{jn}^p e_p\cdot e_i = \Gamma_{jin},
\eeq\noi where $\Gamma_{abc}$ are the Christoffel symbols. On the other hand, these symbols are also given by 
\beq
\Gamma_{jin} &=& \me(\pa_r g_{ij} + \pa_{x^j}g_{in} - \pa_{x^i}g_{jn}) = \me\pa_r(\psi^2 \mr{g}_{ij}) = \psi\psi^\prime\mr{g}_{ij} = \frac{\psi^\prime}{\psi}(\psi^2
\mr{g}_{ij})\n &=& \frac{\psi^\prime}{\psi}{g}_{ij}\eeq\noi where $\mr{g}:= d\Omega\ot d\Omega$ denotes the metric in $S^{n-1}$. So the
 following proposition has just been proved:
\medbreak
\noi {\bf Proposition 1}: $\vvn$ \emph{Let $g: \RR^n\times\RR^n \rightarrow \RR$, given explicitly 
 by $g = dr\ot dr + \psi^2(r)d\Omega\ot d\Omega$,
 be the metric in $\RR^n$, where $d\Omega\ot d\Omega:  T_x(S^{n-1})\times  T_x(S^{n-1}) \rightarrow \RR$ denotes the metric
in $S^{n-1}$. Consider the distance function $r=r(x)$, then for $x = r\Omega, r > 0,$ and $\Omega\in S^{n-1}$ the relation
\bege
Hr(x) = \frac{\psi^\prime(r)}{\psi(r)} (g - dr\ot dr)
\enge is verified, and the Laplacian of $r$ is given by }
\bege
\Delta r(x) = (n-1)\frac{\psi^\prime(r)}{\psi(r)} \vbn
\enge
 \medbreak
 
Suppose now that the metric in $\RR^n$ is given by 
\bege\label{me1}
g = dr\ot dr + \psi_k^2(r) d\Omega\ot d\Omega,
\enge\noi where 
\bege\psi_k(r) = \begin{cases} \frac{1}{\sqrt{-k}} \sinh (\sqrt{-k}r)& k < 0,\\
r& k = 0,\\
\frac{1}{\sqrt{k}} \sin (\sqrt{k}r)& k > 0.
\end{cases}
\enge\noi In the first case ($k < 0$) a metric in AdS is obtained, 
and AdS does correspond to a simply connected hyperbolic spacetime of constant sectional curvature $k$.
It is immediate that 
\bege\label{ma2}
\frac{d\psi_k(r)}{dr} = \begin{cases} \cosh (\sqrt{-k}r)& k < 0,\\
1& k = 0,\\
\cos (\sqrt{k}r)& k > 0.
\end{cases}
\enge\noi From now on we concern the case $k < 0$, describing the AdS geometry. Let $r_k$ the geodesic distance to the origin with respect to the metric (\ref{me1}). 
From eq.(\ref{ma2}), we obtain
\bege
H\ap_k = \sqrt{-k}\;\coth (\sqrt{-k}r)(g-dr\ot dr),
\enge \noi and then it follows that
\bege
\frac{1}{n-1}\Delta r_k
 = \begin{cases} \frac{1}{\sqrt{-k}} \coth (\sqrt{-k}r)& k < 0,\\
1/r& k = 0,\\
\frac{1}{\sqrt{k}} \cot (\sqrt{k}r)& k > 0.
\end{cases}
\enge\noi This result shall still be used. Now we assert and show some properties of the AdS geometry, 
based on the Laplacian Comparison Theorem\footnote{For more details, see \cite{esc}.} 
\medbreak
\noi {\bf Theorem 1}: $\vvn$ \emph{Let $X,Y$ be vector fields in the manifold AdS. 
Assume that the Ricci curvature in all points of the AdS manifold satisfies ${\rm Ric}(X,Y) \geq (n-1)kg.$
If $r$ denotes the geodesic distance to a point $p\in$ AdS and if it is a differentiable function at $x\in$ AdS, then \bege\label{l}
\Delta r(x) \leq \Delta r_k(s),\enge where 
$s = r(x)$. Equality holds only if any sectional curvature along the minimizing geodesic between $p$ and $x$, that contains $r$, is constant and equal to $k$.} $\vbn$
 \medbreak
{\bf Proof}: Let $\ap\in C^3({\rm AdS})$ and $\{e_i\}_{i=1}^n$ be an orthonormal frame field in a tangent space 
at a point $p\in$ AdS. Then 
$||\nabla\ap||^2 = f_if^i$. For each $j = 1,\ldots,n$ it can be verified that 
\bege
\left(\frac{1}{2}||\nabla\ap||^2\right)_j = (\ap^i\ap_i)_j.
\enge\noi Hence, if $\Delta$ denotes the Laplacian, it follows that 
\beq
\frac{1}{2}\Delta(||\nabla\ap||^2) &=& \left(\frac{1}{2}\sum_{j=1}^n||\nabla\ap||^2\right)_{jj}\n
 &=& \sum_{i,j=1}^n (\ap_{ij}\ap_{ij} + \ap_i\ap_{ijj}).
\eeq\noi Since the Ricci equation asserts that $\ap_{ikl} = \ap_{ilk} + R_{kli}^{\;\;\;\;s}f_s$ \cite{ornea}, where $R_{klis}$ denotes the coefficients 
of the Riemann curvature tensor, it follows that
\bege
\frac{1}{2}\Delta(||\nabla\ap||^2)  = \sum_{i,j=1}^n(\ap_{ij}^2 + \ap_i\ap_{jji} + R_{ij}\ap_i\ap_j),
\enge\noi which implies that 
\bege
\frac{1}{2}\Delta(||\nabla\ap||^2) = ||H\ap||^2 + \nabla(\Delta\ap)\cdot \nabla\ap + {\rm Ric}(\nabla\ap,\nabla\ap).
\enge In the particular case when $||\nabla\ap|| = 1$, then  $||H\ap||^2 + (\Delta\ap)^\prime + {\rm Ric}(\nabla\ap,\nabla\ap) = 0$. If 
$r(x) = {\rm distance}\;(x,p)$ is differentiable at $x$ along the minimizing geodesic joining $p$ to $x$, it follows that
\bege\label{ucv}
||Hr||^2 + (\Delta r)^\prime + {\rm Ric}(\pa_r,\pa_r) = 0.
\enge

Now, since $Hr(\pa_r,\pa_r)=0$, the Hessian of $r$ has an eigenvalue equal to zero. Then \bege\label{butc}
||Hr||^2\geq \frac{(\Delta r)^2}{n-1}.
\enge\noi Substituting eq.(\ref{butc}) in eq.(\ref{ucv}), we obtain
\bege\label{ikl}
(\Delta r)^\prime + \frac{(\Delta r)^2}{n-1} + {\rm Ric}(\pa_r,\pa_r) \leq 0.
\enge \noi Using the hypothesis of the theorem, and asserting that ${\rm Ric}(\pa_r,\pa_r)\geq k(n-1)$, eq.(\ref{ikl}) gives
\bege
(\Delta r)^\prime + \frac{(\Delta r)^2}{n-1} + k(n-1) \leq 0.
\enge \noi It is immediate to see that the function 
\bege
\psi = (n-1){\sqrt{|k|}} \coth (\sqrt{-k}r),\quad k < 0,
\enge\noi satisfies the equation $\psi^\prime + \frac{\psi^2}{n-1} + k(n-1) = 0.$

Now, let $\tau:[0,s)\rightarrow \RR$ be a function that satisfies $\tau(0)=0$ and $\psi(\tau(t)) = \Delta r(t)$. But since
\bege\label{p}
(\Delta r)^\prime + \frac{(\Delta r)^2}{n-1} + k(n-1) \leq 0 = \psi^\prime(\tau(t)) + \frac{\psi^2(\tau(t))}{n-1} + k(n-1),
\enge\noi then $(\Delta r)^\prime \leq \psi^\prime(\tau(t))$, which implies 
\bege 
\Delta r^\prime(\tau(t))\tau^\prime(t) \leq \psi^\prime(\tau(t)).
\enge\noi and $\psi^\prime(t) < 0$. Therefore $\tau^\prime(t) \geq 1$, $\tau(t)\geq t$ and
\bege
\Delta r(t) = \psi(\tau(t)) \leq \psi(t),
\enge since $\psi$ is a decreasing function \cite{esc}. If equality holds in eq.(\ref{l}), eq.(\ref{butc}) gives an equality,
and it only happens when $Hr$ has all its eigenvalues equal, except the zero eigenvalue (that comes from $Hr(\pa_r,\pa_r)$).
Since $\Delta r = \Delta r_k$, $\forall r \leq s$, $Hr$ has $n-1$ eigenvectors equal to $\psi/(n-1)$.

Supposing that $\{e_i\}$ diagonalize $Hr$ and that $e_n = \pa_r$, from  eq.(\ref{part}) it follows that
\bege
D_{e_i}\pa_r = \frac{\psi}{n-1}e_i.
\enge\noi At $x$, $[e_i,\pa_r] = 0$, and then
\beq
K(e_i,\pa_r) &=& R(e_i,\pa_r,\pa_r,e_i) = (D_{e_i} D_{\pa_r}\pa_r - D_{\pa_r} D_{e_i} \pa_r - D_{[e_i,\pa_r]}\pa_r)\cdot e_i\n
&=& -\frac{(D_{\pa_r}(\psi e_i))}{n-1}\cdot e_i \n
&=& -\frac{\psi^\prime}{n-1} - \frac{\psi}{n-1}(D_{e_r}e_i)\cdot e_i\n
&=& -\frac{\psi^\prime}{n-1} k - \frac{\psi^2}{(n-1)^2}\n
&=& k, \text{\qquad from eq.(\ref{p})}.
\eeq
\medbreak\hspace{12,5cm} $\square$

\section{Spherical and hyperbolic embedding}

In this section we provide a brief exposition of the Robertson-Walker metric and the hyperbolic embedded coordinates.

\subsection{The Robertson Walker metric}
A 4-sphere has positive curvature and it satisfies the equation 
\bege\label{sph}
\xi_1^2 + \xi_2^2 + \xi_3^2 + \xi_4^2 + \xi_0^2  = R^2,
\enge\noi where $\{\xi_A\}_{A=0}^4$ are Cartesian coordinates in pseudo-Euclidean $\EE^{4,1}$ spacetime and $R$ is the
 4-sphere radius. Then
\bege
d\xi_0 = -\xi_0^{-1}(\xi_1d\xi_1 + \xi_2d\xi_2 + \xi_3d\xi_3 + \xi_4d\xi_4) = -({R^2 - r^2})^{-1/2} ({\bf r}\cdot d{\bf r}),\enge
\noi where ${\bf r} = (\xi_1, \xi_2, \xi_3, \xi_4)$ and $r^2 = {\bf r}\cdot{\bf r}$. On the 4-sphere  the metric is given by 
\beq
g_s &=& d\xi_1\ot d\xi_1 + d\xi_2\ot d\xi_2 + d\xi_3 \ot d\xi_3 + d\xi_4\ot d\xi_4 + d\xi_0\ot d\xi_0\nonumber\\
     &=& dr\ot dr + r^2 d\Omega\ot d\Omega + \frac{1}{R^2/r^2 - 1}dr\ot dr\nonumber\\
     &=& \frac{1}{1-r^2/R^2}dr\ot dr + r^2d\Omega\ot d\Omega,\nonumber\eeq\noi where $d\Omega\ot d\Omega: 
 T_x(S^{3})\times  T_x(S^{3}) \rightarrow \RR $ is the line element on the 3-sphere. If we put the 
time-dependent expansion parameter $a(t)$, the Robertson-Walker metric, written in polar coordinates, is given by
\bege\label{nan}
g = -dt\ot dt + \frac{a^2(t)}{1 - kr^2}dr\ot dr + a^2(t) r^2 (d\theta\ot d\theta + \sin^2\theta d\phi\ot d\phi),\enge\noi where $k = 1/R^2$. If the
 3-sphere is parametrized by polar coordinates
\beq
\xi_4 &=& a\cos\zeta\nonumber\\
\xi_3 &=& a\sin\zeta\cos\theta\nonumber\\
\xi_1 &=& a\sin\zeta\sin\theta\sin\phi\nonumber\\
\xi_2 &=& a\sin\zeta\sin\theta\cos\phi,\qquad 0<\theta,\zeta <\pi,\;\; 0\leq\phi < 2\pi,\nonumber
\eeq\noi  then 
\bege\label{c5}
 d\Omega\ot d\Omega \propto d\zeta\ot d\zeta + \sin^2\zeta (d\theta\ot d\theta + \sin^2\theta d\phi\ot d\phi).
\enge\noi  
Assuming the  map $\sin\zeta dr = r d\zeta$ \cite{6}, it follows the expression 
\bege
\cos\zeta = (1 - kr^2)(1 + kr^2)^{-1}.
\enge\noi  
If it is compared with eq.(\ref{nan}) we see that the metric in eq.(\ref{c5})   
indeed characterizes a spherical manifold. 

\subsection{The hyperbolic embedding}

A 4-hyperboloid has negative curvature and it can be used to describe de Sitter (dS$_{4,1}$, dS$_{3,2}$) or anti-de Sitter
 (AdS$_{1,4}$, AdS$_{2,3}$) spacetimes, according to the  metric signature. We choose to treat the (4,1)-signature case. It satisfies
 the equation \cite{4}
\bege\label{adriana}
\xi_1^2 + \xi_2^2 + \xi_3^2 + \xi_4^2 - \xi_0^2  = R^2,
\enge \noi from which it is immediately seen that  
\bege
d\xi_0 = \xi_0^{-1}(\xi_1d\xi_1 + \xi_2d\xi_2 + \xi_3d\xi_3 + \xi_4 d\xi_4) =  (r^2 - R^2)^{-1/2}({\bf r}\cdot d{\bf r}).\enge
\noi On the 4-hyperboloid surface the metric is given by 
\beq\label{bebe}
g_h &=&  d\xi_1\ot d\xi_1 + d\xi_2\ot d\xi_2 + d\xi_3 \ot d\xi_3 + d\xi_4\ot d\xi_4 - d\xi_0\ot d\xi_0\nonumber\\
     &=& dr\ot dr + r^2 d\Omega\ot d\Omega + \frac{1}{1 - R^2/r^2}dr\ot dr\nonumber\\
     &=& \frac{2 - 1/kr^2}{1-kr^2}dr\ot dr + r^2d\Omega\ot d\Omega,\nonumber\eeq\noi where in this case, $d\Omega\ot d\Omega$ 
denotes the metric on the 3-hyperboloid.

\subsection{The classical model}
 
Schr\"odinger have described  the  reduced model \cite{6}, obtained when one considers the 
 cross section $\xi_3 = \xi_4 = 0$. The $\EE^{4,1}$ spacetime becomes a pseudo-Euclidean (2+1)-spacetime
 endowed with Lorentzian metric
$g  = d\xi_1\ot d\xi_1 + d\xi_2\ot d\xi_2 - d\xi_0\ot d\xi_0$,
 and the 4-hyperboloid given by eq.(\ref{adriana}) becomes 
\bege\label{bu}
\xi_1^2 + \xi_2^2  - \xi_0^2  = R^2.
\enge\noi  If  pseudospherical coordinates are used, in order to parameterize  $\xi_0, \xi_1, \xi_2$, as 
\beq\label{polar}
\xi_1 &=& R\cos\chi\cosh (t/R)\nonumber\\
\xi_2 &=& R\sin\chi\cosh (t/R)\nonumber\\
\xi_0 &=& R\sinh (t/R),\qquad -\infty < t < \infty,\quad 0\leq \chi < 2\pi,
\eeq\noi a map which is nowhere singular is obtained  and it satisfies eq.(\ref{bu}). The metric is then given by  
\bege\label{me}
g_r = -R^2 \cosh^2 d\chi\ot d\chi + R^2 dt\ot dt.
\enge \noi We observe that the new time $t$ varies less rapidly than $\xi_0$.  

In order to recover the full model given by eq.(\ref{bebe}), Schr\"odinger  modified eqs.(\ref{polar}) by two more polar angles, ($\theta, \phi$) and
then the term
\bege \label{en}
d\chi\ot d\chi + \sin^2\chi (d\theta\ot d\theta + \sin^2\theta d\phi\ot d\phi)\enge\noi  is used to enlarge the metric given implicitly by eq.(\ref{me}).
 Besides,
instead of choosing $\chi$ as above, if the relation $\sin\chi = \xi_1/R$ is imposed, another map is defined as follows:
\beq
\xi_1 &=& R\sin \chi\nonumber\\
\xi_2 &=& R\cos\chi\cosh (t/R)\nonumber\\
\xi_0 &=& R\cos\chi\sinh (t/R).
\eeq\noi It reaches another set of pseudopolar angles ($\chi, t$) on the 2-hyperboloid. The metric induced by this parametrization 
is $g_r:= -R^2 d\chi\ot d\chi + R^2 \cos^2\chi dt\ot dt.$ If we want to retrieve the full model again, eq.(\ref{bebe}),  we make the metric 
enlargement
given by eq.(\ref{en}). This is the so-called static frame of the de Sitter metric. In a more familiar form, the
coordinates ($\rho,\eta$) are introduced, parameterizing the reduced de Sitter spacetime as:
\bege
\rho = R\sin\chi, \qquad \eta = Rt, \enge
which gives
\bege\label{quequ}
g_r = - (1 - \rho^2/R^2)^{-1}d\rho\ot d\rho + (1 - \rho^2/R^2)d\eta\ot d\eta.\enge
Formally the de Sitter dS$_{4,1}$ spacetime is uniquely given by the $S^3 \times \RR$ topology of the 4-hyperboloid
\bege\label{on}
\xi_1^2 + \xi_2^2 + \xi_3^2 + \xi_4^2 - \xi_0^2  = R^2
\enge \noi and it can be viewed as a 4-sphere $\xi_1^2 + \xi_2^2 + \xi_3^2 + \xi_4^2 + \xi_0^2  = R^2,$
 if a Wick rotation on the $\xi_0$ coordinate is made: $\xi_0 \mapsto i\xi_0$, where $i$ is the imaginary complex unit. 
If the inferior hemisphere of the Wick-rotated 4-sphere is parametrized, using the coordinates
($t,\rho,\theta,\phi$) as
\beq\label{param}
\xi_0 &=& - R(1-\rho^2/R^2)^{1/2} \cosh(t/R)\n
\xi_1 &=& \rho\sin\theta\cos\phi\n
\xi_2 &=& \rho\sin\theta\sin\phi\n
\xi_3 &=& \rho\cos\theta\n
\xi_4 &=&  R(1-\rho^2/R^2)^{1/2}\sinh(t/R),
\eeq\noi where $ -\infty<t<\infty,\; 0<\rho<R,\; 0<\theta<\pi$ and $0<\phi<2\pi$, the (full model) same metric
given by eq.(\ref{quequ}) is obtained:
\bege\label{eq13}
g = - (1 - \rho^2/R^2)^{-1}d\rho\ot d\rho + (1 - \rho^2/R^2)dt\ot dt - \rho^2 d \theta\ot d\theta 
- \rho^2\sin^2\theta d\phi\ot d\phi.
\enge\noi

\section{Isometry group generators and Killing vectors associated to dS$_{4,1}$}
Before we obtain a description of the Sitter metric which is equivalent to the Schr\"odinger model on Minkowski spacetime,
we discuss how the dS$_{4,1}$ isometry group emerges from the Killing equations related to this spacetime. We also digress
about covariant derivatives and accelerations associated to an arbitrary moving frame on dS$_{4,1}$.

Given the metric (eq.(\ref{eq13}))
\bege
g = - (1 - \rho^2/R^2)^{-1}d\rho\ot d\rho + (1 - \rho^2/R^2)dt\ot dt - \rho^2 d \theta\ot d\theta - \rho^2\sin^2\theta d\phi\ot d\phi,
\enge\noi let us compute the connection Christoffel symbols of dS$_{4,1}$:
\beq
\Gamma^0_{01} &=& \Gamma^0_{10} = -\Gamma^1_{11} = -\rho R^{-2}(1-\rho^2/R^2)^{-1}\nonumber\\
\Gamma^1_{00} &=& \rho R^{-2}(1-\rho^2/R^2),\qquad\qquad\;\;\;\;\;\;\;\;\Gamma^1_{22} = -\rho(1-\rho^2/R^2) \nonumber\\
\Gamma^1_{33} &=& -\rho(1-\rho^2/R^2)\sin\theta,\qquad\qquad\;\;\;\;\Gamma^2_{33} = -\sin\theta\cos\theta\nonumber\\
\Gamma^2_{12} &=& \Gamma^2_{21}=\Gamma^3_{13} =\Gamma^3_{31} = \rho^{-1},\qquad\qquad
\Gamma^3_{23} = \Gamma^3_{32} = \cot\theta\nonumber
\eeq\noi where the index notation was changed: $t\mapsto 0$, $\rho\mapsto 1$, $\theta\mapsto 2$ and $\phi\mapsto 3$.  Using the above 
symbols, the acceleration of moving particles is obtained, which follows immediately from the covariant derivatives. 
First the moving frame
$\{\pa_t, \pa_\rho, \pa_\theta, \pa_\phi\}$ and its respective dual frame $\{dt, d\rho, d\theta, d\phi\}$ is chosen 
in a tangent (cotangent) space at a point on dS$_{4,1}$ spacetime.
 The following expressions 
for the covariant derivative $D$ are obtained:
\beq
D(\pa_\theta) &=& -\rho(1-\rho^2/R^2)d\theta\ot\pa_\rho + \cot\theta\; d\phi\ot\pa_\phi + \rho^{-1} d\rho\ot\pa_\theta
\nonumber\\
D (\pa_t) &=& -\rho R^{-2}(1-\rho^2/R^2)^{-1} d\rho\ot\pa_t -\rho R^{-2}(1-\rho^2/R^2) dt\ot\pa_\rho\nonumber\\
D (\pa_\phi) &=& \cot\theta\; d\theta\ot\pa_\phi -\sin\theta\cos\theta \;d\phi\ot\pa_\theta -\rho(1-\rho^2/R^2)\sin^2\theta\; 
d\phi\ot\pa_\rho +\rho^{-1}
 d\rho\ot\pa_\phi
\nonumber\\
D (\pa_\rho) &=& -\rho R^{-2}(1-\rho^2/R^2)^{-1}(dt\ot\pa_t - d\rho\ot\pa_\rho) +
 \rho^{-1} (d\theta\ot\pa_\theta
+ d\phi\ot\pa_\phi)\nonumber\eeq\noi Applying the covariant derivatives along the orthonormal 
frame vectors, the respective accelerations
are obtained:
\beq
a_\theta &=& D_{\pa_\theta}(\pa_\theta) = -\rho(1-\rho^2/R^2)\pa_\rho\nonumber\\
a_ t &=& D_{\pa_t}(\pa_t) =  -\rho R^{-2}(1-\rho^2/R^2)\pa_\rho\nonumber\\
a_\rho &=& D_{\pa_\rho}(\pa_\rho) =  \rho R^{-2}(1-\rho^2/R^2)^{-1}\pa_\rho\nonumber\\
a_\phi &=& D_{\pa_\phi}(\pa_\phi) = -\sin\theta\cos\theta \pa_\theta -\rho(1-\rho^2/R^2)\sin^2\theta\pa_\rho\nonumber\eeq
\noi Since dS$_{4,1}$ is a maximally isotropic spacetime, it admits a maximal number, ten, of Killing vectors, which can be obtained
immediately from the following Killing equations:
\beq
\pa_\theta\uu_\theta + \om\uu_\rho = 0\n
\pa_\rho\uu_\theta + \pa_\theta\uu_\rho - 2\rho^{-1}\uu_\theta = 0\n
\pa_\rho\uu_\phi + \pa_\phi\uu_\rho - 2\rho^{-1}\uu_\phi = 0\n
\pa_t\uu_\phi + \pa_\phi\uu_t = 0\n
\pa_t\uu_\theta + \pa_\theta\uu_t = 0\n
\pa_t\uu_\rho + \pa_\rho\uu_t + 2R^{-2}\om \uu_t = 0\n
\pa_\rho\uu_\rho - R^{-2}\om^{-1}\uu_t = 0\n
\pa_t\uu_t + R^{-2}\om\uu_\rho = 0\nonumber
\eeq\noi where we introduced $ \om  = \rho(1-\rho^2/R^2)$ and the $\{\uu_\mu\}_{\mu = 0}^3$ evidently denotes the Killing vectors.
It would be desirable to evaluate the Killing equations related to dS$_{4,1}$. 
From the projective embedding to be presented in Sec. 4, the isometries of dS$_{4,1}$ can be obtained from the ones of 
$\EE^{4,1}$. The generators of the isometry group of $\EE^{4,1}$ are given by 
\bege
\uu_A = \Upsilon_{A}^B\xi_B + \si_A,
\enge\noi where $\Upsilon_A^B$ and $\si_A$ ($A, B = 0,1,2,3,4$) are constants. As already pointed, dS$_{4,1}$ is maximally isotropic
and consequently admits a maximal number of Killing vectors, given by (with suitable choices of the $\Upsilon_A^B$): 
\bege
\uu_{AB} = -\xi_A d\xi_B + \xi_B d\xi_A\nonumber
\enge
\noi Using the parametrization given by eqs.(\ref{param}) we obtain explicitly all the isometry generators:

\medbreak
{\scriptsize{\begin{center}
\begin{tabular}{||r||r|r|r|r||}\hline\hline
$p$&$\uu_{0p}$ & $\uu_{1p}$ & $\uu_{2p}$ & $\uu_{3p}$\\\hline\hline
1& $\rho\om\sin\theta\cos\phi\;s$ & $-R\om^{-1}\sin\theta\cos\phi\;c$ & $-\rho R\om\cos\theta\cos\phi\;c$ 
 &$\rho R\om\sin\theta\sin\phi\;c$ \\\hline
2& $-\rho\om\sin\theta\sin\phi\;s$ & $-R\om^{-1}\sin\theta\sin\phi\;c$ & $-\rho R\om\cos\theta\sin\phi\;c$ &
$\rho R\om\sin\theta\cos\phi\;c$ \\\hline
 3&$\rho\om\cos\theta\;c$ & $-R\om^{-1}\cos\theta\;c$ & $\rho R\om \sin\theta \;c$ &0\\\hline
4&$R\om^2$&0&0&0\\\hline
5&0&0&0&$-\rho^2\sin^2\theta$\\\hline
6&0&0&$\rho^2\cos\phi$&$-\rho^2\sin^2\theta\cos\theta\sin\phi$\\\hline
7&$-\rho\om\sin\theta\cos\phi\;c$ & $R\om^{-1}\sin\theta\cos\phi\;s$ & $\rho R\om\cos\theta\cos\phi\;s$ &
$\rho R\om\sin\theta\sin\phi\;s$ \\\hline
8&0&0&$\rho^2\sin\phi$&$\rho^2\sin^2\theta\cos\theta\cos\phi$\\\hline
 9&$-\rho\om\sin\theta\sin\phi\;c$ & $R\om^{-1}\sin\theta\sin\phi\;s$ & $\rho R\om\cos\theta\sin\phi\;s$ &
$\rho R\om\sin\theta\cos\phi\;s$ \\\hline
 10&$-\rho\om\cos\theta\;s$ & $R\om^{-1}\cos\theta\;s$ & $-\rho R\om\sin\theta\;s$ &
0\\ \hline\hline
\end{tabular}\end{center}}\noi Table 1: the Killing vectors associated to dS$_{4,1}$. We assumed the notation $c = \cosh(t/R)$ and $s = \sinh(t/R)$. 
The $p^{\rm th}$ line corresponds to the $\uu_{qp}$
component of the dS$_{4,1}$ isometry group generators ($p = 1,2,\ldots,10$).

\small

\section{Embedded projective coordinates}

The metric given by eq.(\ref{quequ}) can be generalized, 
using embedded projective coordinates. They can be obtained if a stereographic projection
of the 4-sphere on the Minkowski spacetime is done. It is well-known that the Minkowski spacetime 
 can be treated as a tangent space through the 4-sphere South pole. This map
 is the so-called Beltrami (or geodesic) representation \cite{8}. To see how to
 pass from the (4+1)-dimensional formulation to the (3+1)-dimensional orthogonal 
coordinates ($x^0, x^1, x^2, x^3$) on the
Minkowski spacetime we consider the Beltrami representation, which gives the relation \cite{7}

$$x^\mu = - R \frac{\xi^\mu}{\xi^4}.$$\noi Introducing the notation 
\bege\label{pro}
\sigma^2 = x_\mu x^\mu = -x_0^2 + x_1^2 + x_2^2 + x_3^2
\enge\noi  and using eq.(\ref{on})
we have
\bege
\xi^4 = -\frac{R}{(1+\sigma^2/R^2)^{1/2}},  \qquad\qquad  \xi^\mu = \frac{x^\mu}{(1+\sigma^2/R^2)^{1/2}}.
\enge\noi By implicit differentiation it follows that
\bege
d\xi^4 = \frac{x_\mu d x^\mu}{R(1 + \sigma^2/R^2)^{3/2}},  \quad\quad
d\xi^\nu = \frac{(R^2 + \sigma^2) dx^\nu + (x_\mu dx^\mu)x^\nu}{R^2(1+\sigma^2/R^2)^{3/2}}.
\enge\noi Using the above expressions we obtain:
\beq
g= d\xi_A \ot d\xi^A &=& (1 + \sigma^2/R^2)^{-1}dx_\mu \ot dx^\mu\nonumber\\
&&  + R^{-2}(1 + \sigma^2/R^2)^{-2}[x_\mu d x_\mu\ot x^\mu dx^\mu + 2x^\nu x^\mu dx_\nu\ot dx_\mu]\eeq
 \noi This metric is clearly similar to the one given by eq.(\ref{eq13}),  if the 
substitution 
 $\sigma \mapsto i\rho$ is done, obtaining: 
\beq
g &=& (1 - \rho^2/R^2)^{-1}dx_\mu \ot dx^\mu\nonumber\\
&&  + R^{-2}(1 - \rho^2/R^2)^{-2}[x_\mu d x_\mu\ot x^\mu dx^\mu + 2x^\nu x^\mu dx_\nu\ot dx_\mu]\eeq
 \noi Using eq.(\ref{pro}) we can write

\beq\label{sch1}
g &=& (1 - \rho^2/R^2)^{-1}d\rho\ot d\rho\nonumber\\
&&  + R^{-2}(1 - \rho^2/R^2)^{-2}[x_\mu d x_\mu\ot x^\mu dx^\mu + 2x^\nu x^\mu dx_\nu\ot dx_\mu]\eeq
\noi  
As can be seen directly, this expression retrieves the Schr\"odinger description of the static frame of de Sitter metric,
 eq.(\ref{quequ}), or more generally, eq.(\ref{eq13}), if  we consider the metric terms which correspond
 to restriction on Minkowski spacetime.

\small

\section*{Concluding Remarks}
We reviewed the LCT in the light of AdS geometry, describing how the metric in AdS spacetime is related to the 
constant sectional curvature of AdS. Some important features of the topology of AdS are also investigated through 
the proof of LCT an its previous demonstrated proposition. 
From the projective splitting of the de Sitter spacetime on Minkowski one we have expressed hyperbolic (de Sitter) coordinates in
 terms of orthogonal (Minkowski) ones.
The choice of the Beltrami representation seems to be the best one adapted to this formulation. 
Another metric concerning these embedded projective coordinates is constructed, and our formalism retrieves  
 the Schr\"odinger description of the static frame de Sitter metric, if  the terms of the metric restriction
on Minkowski spacetime are considered. 
We have shown that such metric could be obtained from an appropriate parametrization of dS$_{4,1}$, eq.(\ref{param}).
 It would be desirable to obtain the equations of motion, from the Killing
 vectors explicitly evaluated, but it is the purpose of
a forthcoming paper.

%\section*{Acknowledgments}
%One of us (RR) is greatly indebted to  Prof. W A Rodrigues Jr., 
% for stimulating conversations and specially to Dr. R A Mosna for pointing out some mistakes and for giving many other helpful
 %suggestions.

\end{document}